\documentclass[review]{elsarticle}

\usepackage{lineno,hyperref}
\modulolinenumbers[5]

\journal{Journal of \LaTeX\ Templates}

\usepackage{amsmath}
\usepackage[ruled,vlined, linesnumbered]{algorithm2e}

\usepackage{xcolor}

\usepackage{caption}
\usepackage{subcaption}

\let\oldnl\nl% Store \nl in \oldnl
\newcommand{\nonl}{\renewcommand{\nl}{\let\nl\oldnl}}% Remove line number for one line

\begin{document}

\begin{frontmatter}

\title{``How Does It Detect A Malicious App?'' Explaining the Predictions of AI-based Android Malware Detector}

\author[mymainaddress]{Zhi Lu\corref{mycorrespondingauthor}}
\cortext[mycorrespondingauthor]{Corresponding author}
\ead{lu.zhi@stengg.com}

\author[mymainaddress]{Vrizlynn L.L. Thing}
\ead{vrizlynn.thing@stengg.com}

\address[mymainaddress]{Cyber Security Strategic Technology Centre, ST Engineering, Singapore 567710}

\begin{abstract}
AI methods have been proven to yield impressive performance on Android malware detection. 
However, most AI-based methods make predictions of suspicious samples in a black-box manner without transparency on models’ inference.
The expectation on models' explainability and transparency by cyber security and AI practitioners to assure the trustworthiness increases.
In this article, we present a novel model-agnostic explanation method for AI models applied for Android malware detection.
Our proposed method identifies and quantifies the data features relevance to the predictions by two steps: 
i) data perturbation that generates the synthetic data by manipulating features’ values;
and ii) optimization of features attribution values to seek significant changes of prediction scores on the perturbed data with minimal feature values changes.
The proposed method is validated by three experiments.
%First, we show a small proportion of ``relevant'' features identified by our proposed method are helpful in partitioning all the apps by K-means and those from the same malware families have higher probabilities to be in the same clusters.
We firstly demonstrate that our proposed model explanation method can aid in discovering how AI models are evaded by adversarial samples quantitatively.
In the following experiments, we compare the explainability and fidelity of our proposed method with state-of-the-arts, respectively.
\end{abstract}

\begin{keyword}
% \texttt{elsarticle.cls}\sep \LaTeX\sep Elsevier \sep template
% \MSC[2010] 00-01\sep  99-00
Explainable AI \sep Cyber security \sep Machine Learning
\end{keyword}

\end{frontmatter}

%\linenumbers

\section{Introduction}\label{sec:introduction}
\begin{figure*}[tb]
  \centering
  \scalebox{.3}{\includegraphics{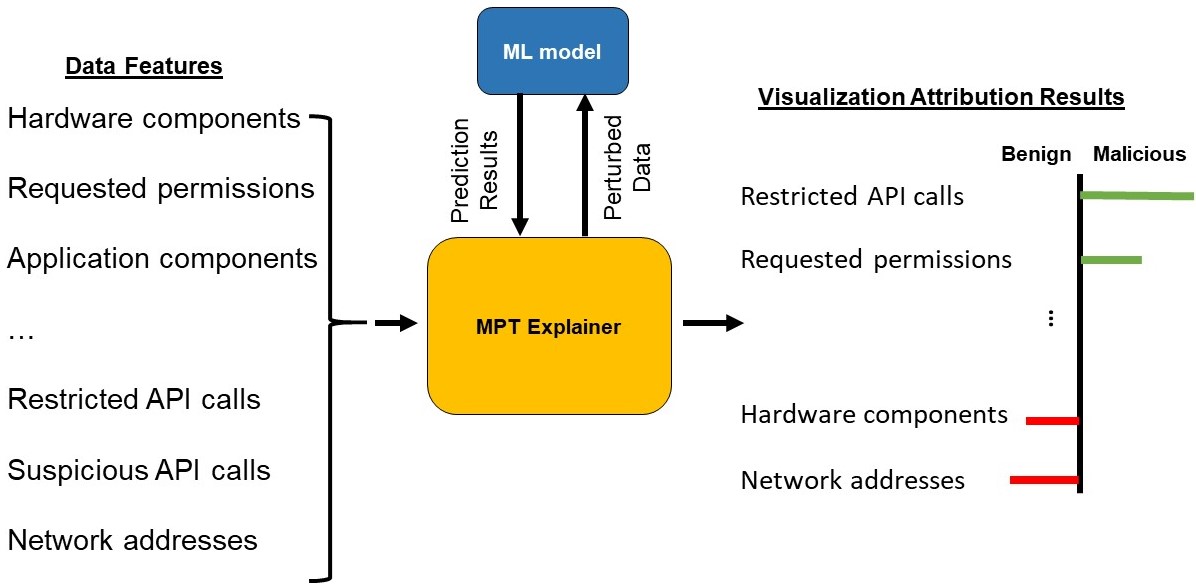}}
  \caption{\textbf{Workflow of explaining an AI model for Android malware detection.} 
   MPT explainer works together with the trained model (blue box) during the model explanation process. Firstly, MPT explainer perturb the input sample by features and observe the model's prediction values. Second, MPT explainer optimizes the features attribution using modern portfolio theory. Finally, each feature's attribution strength is visualized by the bar chart.
  }
  \label{fig_1_work_flow}
\end{figure*}

Artificial intelligence (AI) technologies, in particular shallow machine learning methods (e.g., logistic regression, SVM and random forest, etc.) and deep neural networks (e.g., CNN and LSTM, etc.), have been widely used to fight against cyber attacks today~\cite{buczak2015survey}~\cite{xin2018machine}.
However, explaining the reasoning behind the predictions made by AI models remains one of the most challenging problems in both AI and cyber security communities.
The model explanation is to identify the relevance of data features to the model's predicted class (Fig.~\ref{fig_1_work_flow}).
It is not only to facilitate the cyber security practitioners' understanding on how the AI model identifies the cyber threats, where the confidence on the model's stability is expected, but also helps the AI researchers to find out the models' vulverabilities from certain examples~\cite{eykholt2018robust}. 
The main factors affecting the insights provided by a model explainer are: %:
the challenges of data, such as unbalanced data and high dimentionalities for different applications (e.g., malware detection),
and the model complexity that is about the number of parameters and various types of model structures (e.g, SVM, CNN and LSTM).
These factors have motivated the interest on the \textit{model-agnostic} explainers research~\cite{lime}~\cite{integrated_gradients} that are capable of explaining any types of AI models.

The AI-based malware detection for Android operating system is a well studied problem~\cite{demontis2017yes}~\cite{mclaughlin2017deep}~\cite{yan2018lstm}~\cite{android_malware_LSTM_2019}~\cite{drebin_dataset}.
The popularity and open nature of Android~\cite{idc_os_stats_2020} and its official apps market, i.e., Google Play~\cite{web_google_play_offcial}, have aroused large interests from malware developers, and accordingly make Android users more vulnerable to such attacks by thousands of malicious apps.
In 2019, AV-TEST~\cite{web_av_test_malware_amount_2019} reported 3.16 million Android malware developed.
The efforts on Android malware detection research pay more attention on a secure AI model and the detection performance (e.g., detection rate), where the model explanation is rarely covered.
Demontis et. al.~\cite{demontis2017yes} introduced a secure SVM on Android malware detection that can detect those malware apps evaded from standard SVM-based detectors, through the constraints on the model's parameters selection.
Zhang et. al.~\cite{zhang2019scalable} combined the n-gram analysis and the online classifier techniques to detect the Android malware and attribute their malware families. The online classifier using incremental learning makes it scalable to the rapid evolved malware.
The recently emerging detection methods based on deep neural networks do not only bring astonishing performance in terms of detection rate (i.e., true positive rate), but also remove the needs of features engineering manually, through an automatic joint end-to-end learning~\cite{mclaughlin2017deep}~\cite{yan2018lstm}~\cite{android_malware_LSTM_2019}.
The large amount of parameters and hidden layers structure in deep neural networks, such as CNN and LSTM, make it more difficult for end users to understand the reasoning behind the predictions, even if the authors provided the insights on the effects of performance by the data and network structure manipulation~\cite{android_malware_LSTM_2019}.
However, only recently has the model explanation for malware detection been addressed~\cite{drebin_dataset}.

Clearly, the primary aim of model explanation for Android malware detection is to identify the quantified ``relevance'' between the data features and the predictions.
Existing explainable AI research~\cite{lime}~\cite{integrated_gradients}~\cite{lundberg2017unified} provides model-agnostic explainers for those models that may not be suitable for cyber security issues.
LIME~\cite{lime} explains a model through an optimization~\cite{efron2004least} of a linear model that is human-understandable over the perturbed data with the labels generated by the model's inference.
The high dimensionailty and sparsity of Android malware feature data may cease the explainability of the interpretable models in LIME that is too linear to underfitting. % for understanding purpose.
Integrated Gradients (IGs)~\cite{integrated_gradients} attributes the relevance of data features through the integration of predictions' gradients with respect to the data features varying from zero-values to the real values.
However, the IGs method requires the availability of gradients computation for prediction scores with respect to the model input, which is not available in certain types of classifiers, such as random forests.
The expectation, therefore, on accurate and efficient AI model explanation for cyber security issues, such as malware detection, is still remaining.

In this article, we propose a novel model explanation methodology capable of exploring the reasoning behind the predictions for AI-based Android malware detectors.
The main goal of our methodology is to identify and quantify the relevance of data features with respect to the predictions for any AI-based malware detectors, as compared to the detectors with few human-understandable explanation for specific types of models~\cite{drebin_dataset}.
Furthermore, this methodology explains an AI model by a simple data perturbation technique and efficient linear optimization that is based on the modern portfolio theory (MPT)~\cite{mpt}.
The main advantage of the method based on optimization for features attribution lies in the potential to explain a model accepting a high-dimension data and no requirement on the gradients of prediction scores w.r.t. the inputs.
The proposed methodology can be divided into two stages:
an initial data perturbation technique followed by a linear optimization for feature attribution values (i.e., relevance of features).
The first stage consists of the following steps:
i) data features perturbation by multiplying a scaling factor in $[0,1]$ to have the small changes of feature values (i.e., perturbed data);
and ii) associated prediction scores are obtained by the forward process of the model over these perturbed data.
The second stage then minimizes the \textit{variance} of the multiplication of these prediction scores and the attribution values, with the constraint that the sum of all features' attribution is one.
The \textbf{main contribution} of this article is the formation of the optimization for features attribution, which assumes that the small changes of relevant features affect the prediction scores significantly, and thus are assigned high attribution values.

In addition, results from the quantitative evaluation show that our proposed technique is sensitive to identify the relevant data features to the predictions made by the AI-based Android malware apps detectors, such as SVM~\cite{drebin_dataset}~\cite{zhao2011antimaldroid}~\cite{li2015detecting} and BERT~\cite{devlin2018bert}.
This is verified by the facts that 
1) the feature with the largest attribution value is helpful for the clustering algorithm to group the apps from the same malware family into the same cluster, and more features do not increase the performance significantly;
2) a significant ratio of the camouflaged features activated for the adversarial samples, which evade the malware detection by the classifiers, can be identified.

The rest of the paper is structured as follows:
we present current work on AI-based Android malware apps detection and the model explanation methods in Section~\ref{sec_lit_review}. 
The two learning-based classifiers, i.e., SVM and BERT, are presented in Section~\ref{sec_method_ml}, followed by the introduction to the proposed model-agnostic MPT explainer in Section~\ref{sec_method_mpt}.
We then validate the proposed methodology with the comprehensive experiments in Section~\ref{sec:experiment}. 
Finally, the conclusion of this methodology is discussed in Section~\ref{sec_conclusion}.

% +------------------------------+
% |      Literature Review       |
% +------------------------------+
\section{Literature Review}
\label{sec_lit_review}
In this section, we firstly review the research on Android malware apps detection using AI-based methods, where the early efforts in their work on these models explanation are also covered. Secondly, the general model explanation methods are introduced in terms of their strength and weakness.

\subsection{Android Malware Detection}
The vulnerabilities in Android system are still emerging~\cite{meng2018survey}, and the malware is in rapid evolution.
This requires the development of AI-based models that can predict unknown malware samples in certain degrees.
The AI-based methods described in the Android malware detection literature have focused on four basic features extraction approaches~\cite{bakour2019android}.
In this section, we reviewed two main feature extraction techniques: \textit{the dynamic analysis} and \textit{the static analysis} that are relevant to our work.
The first method for features extraction is \textit{dynamic analysis} that obtain the features via the apps execution, such as system calls and file system usage, etc.
% in a sandbox or real device.
For example, Xiao~\textit{et. al.}~\cite{android_malware_LSTM_2019} propose a deep learning method that analyses the API calls obtained dynamically through two Long-term Short Memory (LSTM) networks trained for malware and non-malware apps, respectively.
 
The second and most commonly used method is the \textit{static analysis technique} that analyses the application source code and resources without execution and requires small running time overhead.
For example, Samra~\textit{et. al.}~\cite{malware_kmeans} use the clustering algorithm in the malware detection that the \textit{permissions} data were extracted from the applications as the features statically.
In~\cite{feizollah2017androdialysis}, a mobile malware analysis tool, AndroDialysis, is developed and used to assess the malware detection performance with \textit{intents} feature (explicit and implicit), which is a run-time binding message object for the inter-process communication in Android framework.
The conclusion drawn from the experiment results shows that intents as the only features in the classification are more effective than that with permissions features.
Zhang~\textit{et. al.}~\cite{zhang2014semantics} convey the semantic information in malware classification via the graph construction of the dependencies among the \textit{API calls} as the features. 
Similarly, Aafer~\textit{et. al.}~\cite{aafer2013droidapiminer} capture the semantic information about the apps' behavior at the API level.
In addition to classification by single features, there are also a number of AI-based methods that classify the malware apps using a combination of features extracted by the static analysis technique.
For instance, Zhu~\textit{et. al.}~\cite{ZHU2018638} detect malware apps by building up an ensemble classifier, Rotation Forests~\cite{rotation_forest} with the features extracted from the APK archive file that includes permissions, sensitive APIs and monitoring system events, etc.
The classifier was trained on a dataset of 2,130 samples of Android apps that malware and benign apps take 50\% respectively, and the results outperforms a SVM classifier.
Arp~\textit{et. al.}~\cite{drebin_dataset} learn a linear SVM classifier with the application features, inlcuding permissions, API calls and network addresses.
They propose a lightweight features extraction method to extract from \textit{mainifest.xml} file and the \textit{dex bytecode}.
% Recently, Xiao~\textit{et. al.}~\cite{android_malware_LSTM_2019} 
In the above methods, only DREBIN~\cite{drebin_dataset} provided a limited explanation of the model's prediction. However, their explanation is based on the weights of the SVM models, which cannot be generalised into other models.

%------------------------------------------------
% Explaining Black-box Android Malware Detection
%------------------------------------------------
Deep learning methods have been successfully applied on the cyber security issues. 
For example, Xiao et. al.~\cite{android_malware_LSTM_2019} proposed a deep learning method that detects the Android malware through training two Long-term Short Memory (LSTM) networks for malware and non-malware apps features (i.e., system calls) respectively.
Vinayakumar~\textit{et. al.}~\cite{vinayakumar2018detecting} tried different LSTM network topologies in Android malware detection.
However, the hidden layers and cells structure in LSTM make the inference process in a black-box mechanism for users, and thus impossible to have the insights on the reasoning.
This reduces the trust on the model's prediction for an example, because it has the risk for the model to be misled to make a wrong decision by an carefully designed example, such as that in the stop sign recognition problem~\cite{eykholt2018robust}, and the user cannot know how the decision is made.

\subsection{AI Model Explanation}
The research efforts in the model explanation literature are mainly to reveal the transparency of the reasoning behind the model's predictions based on the data features, a.k.a., input variables.
There are two major categories of the explanation approaches~\cite{dovsilovic2018_xai_survey}:
i) \textit{transparency-based methods}; and
ii) \textit{post-doc interpretability methods}.
The transparency-based methods~\cite{lipton2018mythos} are the traditional model explanation strategy that conveys the interpretation of the predictions by interpretable AI models. 
That is, these models cannot be too complicated in terms of structure and parameters' numbers such that humans cannot understand~\cite{molnar2019}. 
For instance, in~\cite{drebin_dataset}, the authors tried to enhance the interpretation of the SVM classifier's predictions through the presentation of the features with top ranked weights. Unfortunately, as the complexity of the AI models grows, it is inevitably more difficult for these methods to identify meaningful explanations. In addition, it is trade-off for a model between the transparency for interpretability and the prediction performance~\cite{dovsilovic2018_xai_survey}.

The post-doc methods explored in the literature have raised more attention from both academia and industry.
They extract the causality relationships between the prediction and the data features from the trained (learned) model.
The early research in this field focused on the \textit{model-specific} explanation that seek the rationale of predictions for specific models.
For example, Zeiler~\textit{et. al.}~\cite{zeiler2014visualizing} proposed to visualize and observe the neurons in a convolutional neural networks (CNNs) that shows how each neuron responses to different data instances.
Xu~\textit{et. al.}~\cite{xu2015show} developed an attention-based caption generator for the images, where the attention mechanism shows the highlighted part of the image that is relevant to the generation of a particular word in the caption.
Obviously, in the above methods, it is hard to extend the explanation to other models as the learning-based AI models are rapidly evolving today.
Therefore, there are also several research that aim to develop the model explanation methods that can disclose the inference of predictions for any models, a.k.a.,~\textit{model-agnostic} explanation.
For instance, Samek et. al.~\cite{samek2017explainable} present two methods for deep learning models: (1) sensitivity analysis (SA) that measures how sensitive the predictions are w.r.t. each input variables, and (2) layer-wise relevance propagation (LRP) that propagates the prediction backward in the deep neural network, constrained by a set of propagation rules.
Sundararajan~\textit{et. al.}~\cite{integrated_gradients} propose a feature attribution method, integrated gradients (IGs), that was inspired by a cooperative gaming theory~\cite{aumannvalues}.
It can explain the models that the input data are in different types~\cite{mudrakarta2018did}, such as images, text and tabular data, only if the gradients of model's prediction scores w.r.t. data features is provided.
Besides, several efforts focusing on the complex model explanation that use model approximation. For example, LIME~\cite{lime}, a model-agnostic method for local explanation, explains any classifiers by learning an interpretable model, such as a linear model, with the generated data around the predictions.
However, it has the risk that the linear model for interpretation through approximating the original model make the explanation is not capable of those models and datasets with complicated structure.
Another approach presented by Wu~\textit{et. al.} in~\cite{wu2018beyond} explains the deep models through the original model approximation with a decision tree with few nodes.
The model explanation research efforts also focus on explaining the AI models in different tasks, such as common sense question answering (CQA)~\cite{rajani-etal-2019-explain}.
Recently, Guo et.al.~\cite{guo2018lemna} proposed LEMNA as the solution to LIME's limitations that are caused by the linearity of its surrogate model.
LEMNA is designed specific for the AI models that are used in cyber security tasks, which usually need to deal with sequential data, such as binary code analysis using RNN~\cite{shin2015recognizing}.
The linear approximation of the original model $f(\mathbf{x})$ cannot take the dependencies among features into account that leads to limited capacity of explanation for sequential data.
LEMNA solves this issue using the \textit{fused lasso}~\cite{tibshirani2005sparsity}, which forces the explanation considers the dependencies among features accordingly.
In addition, LEMNA enhanced the fidelity of the local approximation on the non-linearity of the neighborhoods (sampled) around the input data sample $\mathbf{x}$ by \textit{mixture regression model}~\cite{khalili2007variable}.

There are also several development tools for explainable AI (xAI) that are public available online.
The What-If Tool~\cite{wit_google} provides an interactive visual interface for users to inspect the models' behaviors. 
It works with Tensorflow models~\cite{tensorflow2015-whitepaper} as a dashboard that allows users to compare models and find out the importance of features by manipulating the data feature values.
However, this is a tool that requires models running on Google Cloud, and the user-interface is not friendly for those other than data scientists.
Captum~\cite{captum_facebook} is a Python library developed by Facebook recently that implements several model explanation methods, such as DeepLift~\cite{shrikumar2017learning}, Input $\times$ Gradient~\cite{kindermans2016investigating} and Integrated Gradients~\cite{integrated_gradients}.
The functionalities of Captum are under expansion, and this model explanation library currently is only able to explain the deep learning models written in PyTorch~\cite{paszke2017automatic}.
Similarly, SecML~\cite{melis2019secml} provides a Python library that focuses on the secure and explainable AI algorithms.
There are limited number of model explanation methods provided in SecML library, because it is an integrated library that combines several machine learning, adversarial machine learning and explainable machine learning algorithms. 
In addition, SecML only accepts the data instances in the type (i.e., \textit{CArray}) that is defined in the library, which is not a requirement in our proposed method.

% +------------------------------+
% |         Methodology          |
% +------------------------------+
\section{Learning-based Android Malware Detector}
\label{sec_method_ml}
The malware detection is cast into a binary classification problem, i.e., malware and non-malware (benign).
DREBIN~\cite{drebin_dataset}, an Android malware dataset that is used to train and validate the classifiers, is firstly introduced.
Second, the feature representation based on TF-IDF weighting scheme~\cite{manning_raghavan_schutze_2008} that extends the representative of the binary feature vector in~\cite{drebin_dataset} is illustrated.
Finally, we briefly introduce two classifiers: (1) SVM and (2) BERT, that are used in the Android malware detection.

\subsection{DREBIN Dataset}
DREBIN~\cite{drebin_dataset} was developed as a lightweight Android malware detector that utilises the \textit{static analysis technique} to extract the features of suspicious Android apps. An Android malware dataset with the same name, DREBIN, was also released in public.
Each Android application is represented by means of feature sets that were extracted through a linear sweep over the application's manifest file, \textit{AndroidManifest.xml}, and disassembled \textit{dex code} from the bytecode for the Dalvik virtual machine of Android platform.
These feature sets fall into 8 categories, which are numbered from $S_1$ to $S_8$, 
such as $S_2$:~\textit{requested permissions}.
DREBIN datasets collected 5,560 Android malware apps and 123,453 non-malware apps in total.
An Android app is labeled as \textit{malicious} only if it is detected by at least two scanners in VirusTotal~\cite{virus_total}, and \textit{non-malicious} otherwise.

\subsection{Embedding Features Vector by TF-IDF}
The features for an application are in the form of text data, where each feature can be treated as a single ``word'' or token.
It is, thus, necessary to convert the set of features to the vector space that reveals the dependencies among features and the machine learning models can identify the patterns of malicious or non-malicious Apps.
In DREBIN~\cite{drebin_dataset}, the features of an App $\mathbf{X}$ were mapped into a vector space with Boolean values only. 
The mapping function for binary vector values is $\phi: \mathbf{X} \mapsto \{0,1\} \in R^{m}$ that any dimension is marked as $1$ refers to the associated feature contained by the App $\mathbf{X}$, and $0$ otherwise,
where $m$ is the total number of features.
However, the vector representation with Boolean values cannot reveal all the dependencies among features in an App, because of its limited capacity of representation.

We, therefore, extend the representation of a vector with a new mapping function:
\begin{equation}
  \phi(x_i) =
  \left\{
    \begin{matrix}
      \frac{f_{x_i,\mathbf{X}}}{|\mathbf{X}|} \cdot log \frac{|\mathcal{D}|}{|\{\mathbf{X} \in \mathcal{D} : x_i \in \mathbf{X}\}|}, & \text{if feature } x_i \in \mathbf{X} \\
      \\
      0 & \text{, otherwise}
    \end{matrix}
  \right.
\end{equation}

for any feature $x_i$'s numeric representation (i.e., word embedding) in the vector space,
where $|\mathcal{D}|$ is the number of Apps in dataset $\mathcal{D}$,
$|\mathbf{X}|$ is the amount of features in app $\mathbf{X}$,
$f_{x_i,\mathbf{X}}$ is the frequency of $x_i$ in app $\mathbf{X}$,
and $\frac{f_{\mathbf{x}_i,\mathbf{X}}}{|\mathbf{X}|} \cdot log \frac{|\mathcal{D}|}{|\{\mathbf{X} \in \mathcal{D} : \mathbf{x}_i \in \mathbf{X}\}|}$ computes the term weight in TF-IDF weighting scheme.
This weighting scheme guarantees the assignment of relatively low weights on those features appearing frequently among both malicious and non-malicious apps, and relatively high weights on the features that are used in a specific class of apps.
For example, the feature \texttt{android}.\texttt{permission}.\texttt{SEND\_SMS} are more commonly used in malicious apps~\cite{drebin_dataset}.

By the new mapping function $\phi(x_i)$, the vector representation of a malicious app $\mathbf{X}$ in the family \texttt{Plankton} may look like:

{
  \[
  \left (
    \begin{matrix}
      0\\
      0.13\\
      \textbf{...}\\
      0.08
    \end{matrix}
  \right )
    \begin{matrix}
      \texttt{m}.\texttt{facebook}.\texttt{com}\\
      \texttt{android}.\texttt{hardware}.\texttt{wifi}\\
      \textbf{...}
      \\
      \texttt{android}.\texttt{hardware}.\texttt{touchscreen}
      
    \end{matrix}
    \begin{matrix}
      &S_8\\
      &S_1\\
      \\
      &S_1  
    \end{matrix}
\]
\centering
}

\subsection{Learning-based Malware App Detection}
We cast the malware detection problem as a binary classification task, where the \textit{malware} is labeled as the positive class and the \textit{benign} app (\textit{non-malware}) is the negative class.
Specifically, two popular classifiers, i.e., SVM and BERT, are trained to separate the malicious apps from the non-malicious (benign).

\textbf{SVM} is a classical and popular classifier for binary classification before the deep learning era.
It maps a sample into a high-dimension space that has a clear gap between two classes, and thus makes the prediction of the sample's category more accurate.
It has been proven effective on Android malware detection by many early research~\cite{drebin_dataset}~\cite{li2015detecting}~\cite{zhao2011antimaldroid}.
We train a SVM model with Radial basis function (RBF) kernel~\cite{vert2004primer}.
Each feature of the apps are taken as a token in text analysis, and vectorized by TF-IDF weighting scheme.
The trained SVM produces the prediction scores for each class, which indicates its confidence on the prediction for a class.
In addition, the state-of-the-art deep learning method for natural language processing (NLP), \textbf{BERT}, is employed as another classifier.
In order to have the trade-off between the performance and the hardware resources, we train the base uncased BERT model implemented by Huggingface~\cite{Wolf2019HuggingFacesTS} with a maximum length of input text as 128.
Both classifiers are trained on the balanced DREBIN dataset.
The detection performance shown in Table~\ref{tab:svm_bert_classifier_perf_original} denote to the high detection rates (0.9684 for SVM and 0.9591 for BERT) with low false positive rates (0.0425 for SVM and 0.0420 for BERT).

\section{Evade Detection by Adversarial Samples}
\label{sec_adv_samples}
A machine learning model for malware detection can be vulnerable to attacks using well-crafted adversarial samples, because of the intrinsic assumption to build up the model that the training and test data are drawn from the same distribution~\cite{demontis2017yes}~\cite{barreno2010security}.
The adversarial sample violates such assumption by activating a certain number of features in the features vector to make it available to evade the AI-based malware detection without compromising of its malicious functionalities.
% Attackers may evade the AI-based malware detection using adversarial samples, which features are carefully crafted in order to camouflage the malware as the benign apps without compromising of the malicious functionalities.

\subsection{Adversarial Samples Generation Algorithm}
In this paper, we simulate an evasion attack using the adversarial samples to mislead SVM and BERT classifiers in malware detection.
The attacker is assumed to have full capacity of manipulating the data in the inference stage, including the knowledge of the feature space, and the model's feedback, such as prediction scores.
The adversarial samples are generated with the following settings:
Firstly, the optimised set of features for the adversarial samples are selected/activated through a process of heuristic search that maximises the fitness function~\cite{liu2019adversarial}.
Then, the manipulated samples will be converted into the features space by calculating the TF-IDF values for the features in the adversarial samples, including the original and the activated features.
Please note that we only activate the features under the ``Permission'' category, in order to preserve the malicious functionalities, as suggested as by~\cite{demontis2017yes}.
We also extend the genetic algorithm~\cite{liu2019adversarial} that was used in the adversarial samples generation, where the model's prediction score of benign class is used as the fitness values.
This guarantees that (1) the adversarial samples can evade the detection, because of the prediction scores $>$ 0.5 for benign class; (2) we have the quantified confidence that the classifier incorrectly classifies them into benign class, which will be used in later experiments (see Section~\ref{sec:exp:adv_samples_features_analysis}).
The pseudo code for the adversarial samples generation is illustrated in Algorithm~\ref{alg:the_alg_1}.
The MPT explainer is then used to explain why the classifiers are bypassed by these adversarial samples, where the activated features in the adversarial samples are expected to be attributed with high values.
% That is, it is expected for MPT explainer to attribute the activated features in the adversarial samples with high values.

% \subsection{Generation of Adversarial Samples}
Specifically, we run the genetic algorithm to generate the adversarial samples dataset by the following parameters:
(1) there are 200 malware samples randomly selected from the test set that will be camouflaged to fool the classifiers in the experiments, which are called \textbf{base samples};
(2) the maximum iterations for the evolution of solutions in genetic algorithm are 500 loops, which is based on our observation in the early experiments that is enough to guarantee the adversarial samples can be generated for most base samples.
In addition, the optimisation process of the genetic algorithm will be converged if any of the following conditions is fulfilled:
(1) the maximum prediction score of the adversarial samples has been idle for at least 10 iterations that implies the no solutions (i.e., a set of activated features) with higher fitness values can be obtained through the optimisation process in reasonable time;
(2) the maximum fitness value $> 0.99$, i.e., the prediction score, which means the classifiers cannot distinguish the adversarial samples from the malware samples at all, and there is little benefits to increase the fitness value by the genetic algorithm;
or (3) the solutions in the optimisation process have been evolved for 500 loops.

By these settings, we generated $29,821$ adversarial samples that can evade both SVM and BERT classifiers.% as the candidates pool. 
% However, the samples distribution by their prediction scores are not balanced
The samples are distributed by their model's prediction scores, where the interval is $0.1$ and most of the adversarial samples evade the model's prediction (SVM and BERT) with scores of $\leq 0.9$.
%, which is shown in Figure~\ref{fig:adv_samples_distribution}.
We randomly pick up 100 samples for each prediction scores interval from the pool (i.e., totally, 500 samples for BERT and 499 for SVM that one sample is removed because it was false negative in the original prediction, respectively) for both SVM and BERT classifiers for a fair assessment on the model explanation experiment. The MPT explainer is then used to attribute the features for each adversarial sample such that the ``relevant'' features to the false negative predictions can be identified.
More details about the explanation on the adversarial samples by MPT explainer will be discussed in Section~\ref{sec:experiment}.

\begin{algorithm}

\SetAlgoLined
\SetKwInOut{Input}{Input}
\SetKwInOut{Output}{Output}
\Input{
A malware sample from DREBIN dataset.
\newline
SVM classifier
}
\Output{A set of adversarial samples.}

\nonl \textbf{Stage 0 - Candidate Features}

Collect features from training set under ``Permission'' category.

Exclude features in the input malware sample.

\nonl \textbf{Stage 1 - Construction of Adversarial Samples}

Initialize a ``population'' matrix that random features are activated in each solution.

\nonl \textit{shape=(num\_solutions, num\_genes)}.

\nonl One \textbf{solutions} refers to an adversarial sample candidate.

\nonl The \textbf{genes} means the candidate camouflaged features.

iter\_counter = 0

MAX\_LOOP = 500

\While{True}{

Compute fitness values - \textbf{p}

Select matting pool

Crossover

Mutation

Update ``population''

iter += 1

\uIf{iter\_counter $\geq$ MAX\_LOOP}{
\textbf{break}
}

\uIf{max(\textbf{p}) idle for 10 iterations}{
\textbf{break}
}

\uIf{max(\textbf{p}) $>$ 0.99}{
\textbf{break}
}
}

\Return Adversarial samples that fitness values $>$ 0.5 
\newline
\textbf{and} prediction score of benign class by BERT $>$ 0.5

\caption{Adversarial Samples Generation}
\label{alg:the_alg_1}
\end{algorithm}

% +------------------------------+
% |         Methodology          |
% +------------------------------+
\section{Explaining Classifier's Predictions}
\label{sec_method_mpt}
In this section, we firstly formulate the model explanation as a feature attribution problem that is to quantify the relevance of features to the classifier's prediction.
The data perturbation technique, then, for the generation of the candidate data in model explanation is introduced.
Finally, we present the details of the proposed methodology, i.e., MPT Explainer, that is based on modern portfolio theory (MPT)~\cite{mpt}.

\subsection{Theory Basis}
{\bf Modern Portfolio Theory} (MPT) is an economic theory that maximize the expected returns through a portfolio of assets (e.g., an allocation of investment on a set of stocks), given a certain level of risk.
Inspired by MPT, the features in a data sample can be treated as the assets. The output attribution values for the features, which refer to the explanation, are similar with the allocation of the investment, given a certain level of the perturbation of the classifier's prediction scores.

\subsection{Problem Statement}
{\bf Problem statement}
The classifier $f(\mathbf{x}) :\to [0,1]^{|C|} \in \mathcal{F}$ is a mapping function to map the data instance $\mathbf{x}$ to a set of probabilities,
which denote to the confidence of the classifier $f$ to classify $\mathbf{x}$ to each class.
A data instance $\mathbf{x}=(x_1, x_2, ..., x_m) \in R^m$ is a vector with $m$ features that a feature $x_i$ can be the numerical representation of either a token in text data or a pixel (or a superpixel) in image data.
The \textbf{explanation} of a classifier $f$ is to seek the quantified relevance of each feature $x_i$ in data instance $x$ to the predicted class. % by classifier $f$.
Specifically, this is called \textit{feature attribution} that is in the form of an attribution vector $\mathbf{A} = (a_1, a_2, ..., a_m) \in R^m$, 
where $a_i \in [-1,1]$ represents a degree of relevance of feature $x_i$ to the prediction, and the lower the less relevant.

\subsection{Data Perturbation}
We look for the interpretation of the classifier's prediction with respect to single data instances, which is also known as \textit{local interpretation}~\cite{lime}.
The challenge of local interpretation is the limited number of data that can be used to attribute the features' relevance, and the observation of the changes of prediction scores by different data becomes impossible.
Therefore, the synthetic data generation through perturbing the values of each feature becomes necessary for further model explanation.
Specifically, given a data instance $\mathbf{x}=(x_1, x_2, ..., x_m)$ from the dataset, the classifier outputs a prediction $f(\mathbf{x})$ that is the probability of the predicted class (e.g, in binary classification, this is the larger probability).
For each feature $x_i$ ($i = {1,2,3,...m}$), the $K$ perturbed data over this feature are computed as:

\begin{equation}
\label{eqn:data_perturbation}
    x_i^{(k)} = \alpha_i^{(k)} \cdot x_i
\end{equation}

\noindent where $k \in \{1,2,3, ..., K\}$ is the number of perturbed data, and $\{\alpha_i\}\in[0,1]$ are evenly spaced and the coefficients to linearly change the feature \textit{i} values slightly. Based on our observation in the early experiments, $K$ is set as 50 for the trade-off between capacity of data perturbation and computation efficiency.

By the data perturbation strategy, we simply obtain a set of $K \times |\mathbf{x}|$ perturbed data, which is $\mathbf{P_x}$.
These data will be fed into the classifier $f$ to have the prediction scores over the classes.
Both the perturbed data and the associated prediction scores will be used in the optimization of the attribution values vector $\mathbf{A}$ (See Section~\ref{sec:opti_mpt}).

\subsection{Optimization for Features Attribution}
\label{sec:opti_mpt}
The purpose of model explanation is to find the optimized attributions values for the features.
The assumption behind is that the prediction scores are more sensitive to the changes of relevant features than those relatively less relevant.
That is, a small changes of relevant features values will cause significant changes of prediction scores.
With the perturbed dataset $\mathbf{P_x}$, the changes of prediction scores between the $\mathbf{x^{(ik)}}$ (i.e., the k-th perturbed data instance by perturbing the i-th feature value) and the original data instance $\mathbf{x}$ is defined as the fraction of the prediction scores difference between $f(\mathbf{x^{(ik)}})$ and $f(\mathbf{x})$:

\begin{equation}
    r_i^{(k)} = \frac{f(\mathbf{x^{(ik)})} - f(\mathbf{x})}{f(\mathbf{x})}
\end{equation}

This forms a $K$-dimension vector of the prediction scores changes, $\mathbf{r_i} = (r_i^{(1)}, r_i^{(i2)}, ..., r_i^{(K)})$, when varying the values of feature $x_i$ by equation~(\ref{eqn:data_perturbation}).

We model the association between the attribution values of features, i.e., $\mathbf{A}$, with the random prediction scores changes, $\mathbf{r}=(r_1, r_2, ..., r_m)$, for all $m$ features by their inner product:

\begin{equation}
    \mathbf{A^T}\mathbf{r} = \sum_{i=1}^{m}{r_i a_i}
\end{equation}

we compute the variance of this association as:

\begin{equation}
    \begin{split}
        Var[\sum_{i=1}^{m}{r_i a_i}] & = E[(\sum_{i=1}^{m}{r_i a_i} - E(\sum_{i=1}^{m}{r_i a_i}))^2] \\
        &= E[(\sum_{i=1}^{m}{r_i a_i} - a_i\mu_i)(\sum_{j=1}^{m}{r_j a_j} - a_j\mu_j)] \\
        &= \sum_{i=1}^{m}\sum_{j=1}^{m}{a_i a_j}\cdot E[(r_i - \mu_i)(r_j - \mu_j)] \\
        &= \sum_{i=1}^{m}\sum_{j=1}^{m}{a_i a_j \sigma_{ij}} = \mathbf{A^T}\mathbf{Q}\mathbf{A}
    \end{split}
\end{equation}

\noindent where $\mathbf{Q}$ is the covariance matrix.

The aim is to find out an attribution vector, $\mathbf{A}$, that the variance of this association $\mathbf{A^T}\mathbf{r}$ is minimized, such that the non-trivial changes of the prediction scores can be obtained with the minimal fluctuations of the data features.
The function of the optimization is given by:

\begin{equation}
\label{eqn:mpt_optimization}
    \begin{split}
        &\arg\min_{\mathbf{A}}{\mathbf{A^T}\mathbf{Q}\mathbf{A}} \\
        & \textit{w.r.t.} ~\sum_{i=1}^{m}{a_i}=1~\text{and}~ a_i\in[-1,1]
    \end{split}
\end{equation}

The equation (\ref{eqn:mpt_optimization}) considers both the individual features' contribution to the final predictions and the contribution made by the dependency among these features.
Note that we argue that the features with negative attribution values do not mean that they have negative contribution to the prediction. 
The negative value means, instead, the associated feature has negligible affect on the final prediction.

% +------------------------------+
% |         Experiments          |
% +------------------------------+
\section{Experiments}
\label{sec:experiment}
\subsection{Experiment setup}
The dataset utilised in this paper consists of 11,110 samples of Android apps drawn from DREBIN dataset, in which 
% The benign and malware apps each account for half of the dataset (i.e., 5,555 benign apps and 5,555 malware apps).
there are $5,555$ benign apps and $5,555$ malware apps, respectively.
Further, we randomly separate this dataset into two dis-joint sub-sets: a training set with 7,442 apps and a test set with 3,668 apps.
There are 179 malware families and one benign family in the whole dataset that the top four malware families (i.e., FakeInstaller ($15.7\%$), Plankton ($11.8\%$), DroidKungFu ($11.7\%$), Opfake ($11.0\%$)) with the largest number of samples account for more than $50\%$ malware samples
%including Plankton, BaseBridge and FakeDoc, etc..
The apps in the training set are from 160 malware families and the test set covers 124 malware families, respectively.
% Fig.~\ref{fig:apps_ratio_by_family} shows the distribution of malware families in the test set that the four largest malware families (i.e., FakeInstaller, Plankton, DroidKungFu, Opfake) account for more than 50\% malware families.
% \begin{figure}[t]
%     \centering
%     \scalebox{.55}{\includegraphics{./fig_pie_charts_pie_malware_apps_ratio_with_family_names_BERT_2020-03-17.png}}
%     \vspace*{-15mm}
%     \caption{\textbf{Distribution of apps amount by malware family in test set.} The four largest categories of malware families (i.e., FakeInstaller, Plankton, DroidKungFu, Opfake) account for more than 50\% of the total number of malware families.}
%     \label{fig:apps_ratio_by_family}
    
% \end{figure}

The two classifiers, i.e., SVM and BERT, are trained over the training set with the following settings:
(1) \textbf{SVM} is with RBF kernel that gamma = $1.0$.
(2) \textbf{BERT.} We use the uncased base model implementation of BERT from HuggingFace Transformers library~\cite{Wolf2019HuggingFacesTS} with the default configurations of parameters, e.g., the maximum sequence length of the text is 128, learning rate is 4e-5. 
The batch size for training is set up as 8 for a trade-off between the performance and the memory limitation, and the model was trained for 5 epochs.
Since BERT uses the word embedding technique to generate numeric representation of the text tokens that is not easy to perform the data perturbation directly by our method, we approximate its prediction behaviour by training a SVM, called \textit{aprox\_SVM}. Specifically, BERT predicts a set of training data to have the predicted labels. The aprox\_SVM was trained on the same dataset and the ``ground truth'' are the predicted labels by BERT. By this way, the aprox\_SVM has a TPR=0.9984 and FPR=0.0029 that means it has a highly similar performance as the BERT.
The aprox\_SVM will only be used to generate the predictions of perturbed data in the model explanation.
Note that in the experiments, the features attribution for BERT is identified by aprox\_SVM with MPT explainer. 

The classifiers' performance were assessed on the test set and summarized in Table~\ref{tab:svm_bert_classifier_perf_original} with true positive rate (TPR), false positive rate (FPR) and Matthews correlation coefficient (MCC). 
The MCC combines true/false positives/negatives into a single metric range from $[-1,1]$ that 1 means a perfect prediction, 0 denotes to the prediction performance that is no better than the random prediction and -1 indicates a total disagreement between the prediction and ground truth.
It shows that the predictions by both classifiers match the ground truth quite well in terms of MCC, and SVM has a slightly better performance than BERT on Android malware detection, although SVM has approximate 0.0005 higher of FPR.

\begin{table}[h]
\caption{\textbf{Classification performance for SVM and BERT.} Both classifiers obtained high true positive rate (TPR) and MCC, and relatively low false positive rate (FPR). It shows a high detection rate for both classifiers in the Android malware detection task. }
\centering
\label{tab:svm_bert_classifier_perf_original}
\begin{tabular}{|c|c|c|}
\hline
                                                                                           & \textbf{SVM}    & \textbf{BERT}   \\ \hline
\textbf{True Positive Rate (TPR)}                                                          & \textbf{0.9684} & 0.9591          \\ \hline
\textbf{False Positive Rate (FPR)}                                                         & 0.0425          & \textbf{0.0420} \\ \hline
\textbf{\begin{tabular}[c]{@{}c@{}}Matthews correlation \\ coefficient (MCC)\end{tabular}} & \textbf{0.9259} & 0.9171          \\ \hline
\end{tabular}
\end{table}

The parameter of the perturbed data amount around the input data instance $\mathbf{x}$ is set as 50 for the proposed MPT Explainer.
For example, for an data instance $\mathbf{x}$ with 10 features to describe the suspicious app, there are 500 perturbed data instances generated where each feature $x_i$ was varied into 50 values.

Next, we present three quantitative experiments to assess the MPT Explainer's capability of model explanation. 
%The first experiment (see Section~\ref{sec:exp:kmeans}) is to validate the efficacy of the features with $N\in[1,10]$ largest positive attribution values through the K-means clustering on suspicious apps.
In the first experiment (see Section~\ref{sec:exp:adv_samples_features_analysis}), we test the capacity of features attribution by MPT explainer for adversarial samples that can evade the detection by the classifiers, i.e., SVM and BERT.
We also compare the MPT explainer with another two state-of-the-art explainers (see Section~\ref{sec:exp:good_comparison}), which are LIME~\cite{lime} and SHAP~\cite{lundberg2017unified}, in the second experiment of ``good'' explanation.
Finally, a fidelity test (Section~\ref{sec:exp:fidelity_test}) on the explanation is conducted for MPT explainer and SHAP.

\subsection{Analysis of Activated Features in Adversarial Samples}
\label{sec:exp:adv_samples_features_analysis}

In this experiment, we evaluate if the MPT explainer is able to identify the activated features in an adversarial example (see Section~\ref{sec_adv_samples}).
It is expected that the MPT explainer can help to find out the reasons that the adversarial example evades the model's detection in the features level.
Specifically, we evaluate the explanation performance in two ways: (1) the general performance assessment, and (2) the functional analysis in terms of the activated features number in an adversarial sample and the prediction scores, respectively.

The \textbf{first assessment} is to observe the general performance of MPT explainer on features attribution based on the percentage of the samples with ``good'' explanation.
A sample has \textbf{``good'' explanations}, only if a certain percentage of its activated features (i.e., camouflaged features) are attributed with positive values.
The threshold for a sample to have ``good'' explanation varies from 0\% to 90\% with 10\% intervals in our experiment.
The increased thresholds of ``good'' explanation means less tolerance for MPT explainer to incorrectly attribute the activated features with negative values.
In Table~\ref{tab:general_performance_explanation}, it is clear that the amount of adversarial samples with “good” explanation by MPT Explainer is decreasing as the “good” explanation threshold is increasing. 
Specifically, in the \textbf{Configuration 1} that candidate activated features are from benign apps samples in the training set, the MPT explainer identifies a large number of the activated adversarial features as the positive contribution to the false negative prediction of benign class, when the “good” explanation threshold is less than 50\%.
In the usage of an analysis tool, such explanation of the classifier’s behavior on adversarial samples can reveal how the detector was evaded.
The explanation performance, however, is deteriorated quickly once the “good” explanation threshold is larger than 50\%.
We argue that such performance deterioration is reasonable, because the model's prediction for a sample is made by the cohesive contribution of all the features.
%the classifier is based on the calculation over the sample’s features vector.
That refers to a comprehensive consideration of all features by the classifier (i.e., over the distribution of the features vector), including the contributions made by individual features and that made by the complicated relationships among features.

Further, we verified the above argument by another experiment with the \textbf{Configuration 2} that the candidate activated features are from \textit{malware samples} in the training set.
In Table~\ref{tab:general_performance_explanation}, it shows quite high percentages of adversarial samples with ``good'' explanation for both SVM and BERT, when the thresholds of ``good'' explanation are larger than 50\%. 
Note that the activated features were used by malware samples and intuitively, all the features in malware samples should have been no contribution to the false negative predictions of benign class. 
Therefore, the features are attributed with positive values by MPT explainer imply that the classifiers consider all the features comprehensively.
That is, a malware sample is camouflaged as a benign app sample to evade the detection successfully by:
(1) the contribution from individual activated features;
and (2) the contribution made by the activated feature and the existing features together, which is why some of the features in the original base sample are also attributed with positive values.

\begin{table*}
\centering
% \captionsetup{labelformat=empty}
\caption{\textbf{Results on adversarial samples explanation in terms of the percentage of ``good'' explanation}. The varying thresholds of ``good'' explanation (top row) are from $>$ 0\% to $>$90\% with 10\% intervals. \textbf{C} for configuration. \textbf{S} for SVM and \textbf{B} for BERT.}
\begin{tabular}{|c|c|c|c|c|c|c|c|c|c|c|c|} 
\hline
~                    & ~               & \textbf{ 0\% } & \textbf{10\%} & \textbf{20\%} & \textbf{30\%} & \textbf{40\%} & \textbf{50\%} & \textbf{60\%} & \textbf{70\%} & \textbf{80\%} & \textbf{90\%}  \\ 
\hline
\textbf{ C1 } & \textbf{S}  & 1                            & 1                             & 0.992                         & 0.908                         & 0.581                         & 0.309                         & 0.216                         & 0.188                         & 0.188                         & 0.182                          \\ 
\hline
\textbf{ C1 } & \textbf{B} & 1                            & 1                             & 0.994                         & 0.882                         & 0.592                         & 0.336                         & 0.23                          & 0.196                         & 0.188                         & 0.18                           \\ 
\hline
\textbf{ C2 } & \textbf{S}  & 1                            & 1                             & 1                             & 0.99                          & 0.936                         & 0.679                         & 0.293                         & 0.164                         & 0.142                         & 0.128                          \\ 
\hline
\textbf{ C2 } & \textbf{B} & 0.996                        & 0.996                         & 0.996                         & 0.994                         & 0.948                         & 0.640                         & 0.348                         & 0.232                         & 0.204                         & 0.194                          \\
\hline
\end{tabular}
\label{tab:general_performance_explanation}
\end{table*}

% \begin{figure*}
%     \centering
%     \scalebox{.40}{\includegraphics{./fig_4_adv_samples_distribution.png}}
%     \caption{\textbf{Adversarial Samples Distribution by Prediction Scores.} The amount of samples for different prediction scores are extremely unbalanced, which cannot be used directly in the experiments for a fair assessment.}
%     \label{fig:adv_samples_distribution}
% \end{figure*}

The \textbf{second assessment} is a \textit{functional analysis} that measures the adversarial samples with ``good'' explanation, as a function of two variables: 
(1) the prediction score of benign class by the classifier;
and (2) the number of activated features in each sample.
A higher prediction score indicates the stronger confidence from the classifier on its prediction, when the classifier considers all the features, including the contributions made by individual features.
In addition, more features are activated in an adversarial sample denotes to the significant changes of the features vector, which implies more complicated relationships among features, and thus the features attribution are more difficult than that on the adversarial samples with less amount of activated features.

\begin{table}
\caption{\textbf{Function analysis for model explanation (SVM left, BERT right), as a function of the model's prediction scores and the number of activated features in each sample.} The threshold of an adversarial sample with the “good” explanation $>$ 0.4. The candidate activated features are by Configuration 1. The intervals of the classifier's prediction scores are shown in the first column, and the intervals of the number of activated features in a sample is shown in the first row.}

\begin{tabular}{|l|l|l|l|} 
\hline
      SVM               & \textbf{			(0, 30]		} & \textbf{			(30,50]		} & \textbf{			50+		}  \\ 
\hline
\textbf{			(0.5,0.6]		} & 0.448                 & 0.278                 & 0.286              \\ 
\hline
\textbf{			(0.6,0.7]		} & 0.392                 & 0.500                 & 0.111              \\ 
\hline
\textbf{			(0.7,0.8]		} & 0.661                 & 0.636                 & 0.185              \\ 
\hline
\textbf{			(0.8,0.9]		} & 0.737                 & 0.636                 & 0.490              \\ 
\hline
\textbf{			(0.9,1.0]		} & 1.000                 & 1.000                 & 1.000              \\
\hline
\end{tabular}
\quad
\begin{tabular}{|l|l|l|l|} 
\hline
       BERT             & \textbf{			(0, 30]		} & \textbf{			(30,50]		} & \textbf{			50+		}  \\ 
\hline
\textbf{			(0.5,0.6]		} & 0.579                 & 0.333                 & 0.417              \\ 
\hline
\textbf{			(0.6,0.7]		} & 0.444                 & 0.545                 & 0.706              \\ 
\hline
\textbf{			(0.7,0.8]		} & 0.600                 & 0.167                 & 0.786              \\ 
\hline
\textbf{			(0.8,0.9]		} & 0.553                 & 0.222                 & 0.667              \\ 
\hline
\textbf{			(0.9,1.0]		} & 0.667                 & 0.667                 & 0.909              \\
\hline
\end{tabular}
\label{tab:func_analysis_svm}
\end{table}

We show the analysis results for SVM and BERT in Table~\ref{tab:func_analysis_svm}.
The number in each cell is calculated by the number of adversarial samples with ``good'' explanation divides the number of the adversarial samples that has the prediction scores in a certain interval (i.e., a row in the table) and the amount of activated features (i.e., a column in the table).
The functional analysis reveals the features attributions for both SVM and BERT by MPT explainer have better performance when the classifier has a higher confidence on its false negative prediction of benign class (i.e., prediction score close to 1) than those sample with less confidence.
In addition, the performance of features attribution by MPT explainer for SVM is deteriorated if the number of activated features in an adversarial sample is increased.
And more activated features in the adversarial samples are identified by the MPT explainer for BERT as the number of activated features in a sample increased.
The difference reveals the following findings: 
(1) SVM and BERT made the false negative prediction by considering different features; 
and (2) it requires to activate less amount of features to evade the detection by SVM than that by BERT, which maybe helpful for the analysts to find out the solution to a more secure malware detector.

\subsection{Explanation Capability Comparison}
\label{sec:exp:good_comparison}
\begin{figure}
    \centering
    \begin{subfigure}[b]{0.4\textwidth}
      \centering
      \includegraphics[width=\textwidth]{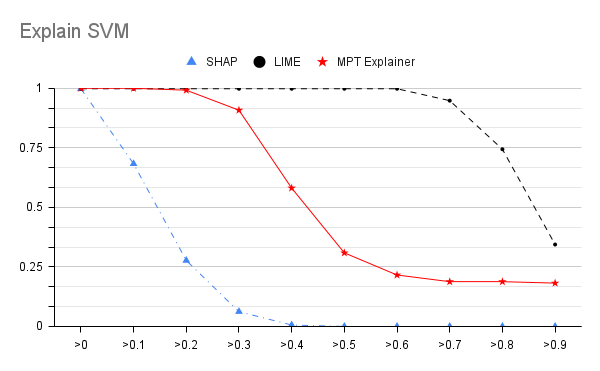}
      \caption{Percentage of ``good'' explanation identifid by LIME, SHAP \& MPT Explainer for SVM}
      \label{fig:compare_lms_good_explain_a}
    \end{subfigure}
    \hfill
    \hfill
    \begin{subfigure}[b]{0.4\textwidth}
      \centering
      \includegraphics[width=\textwidth]{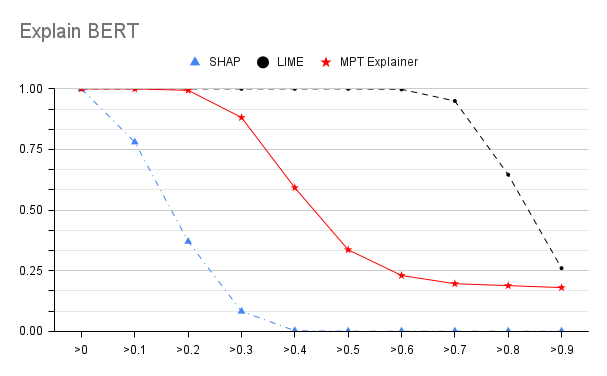}
      \caption{Percentage of ``good'' explanation identifid by LIME, SHAP \& MPT Explainer for BERT}
      \label{fig:compare_lms_good_explain_b}
    \end{subfigure}
    \caption{\textbf{``Good'' segmentation comparesion}}
    \label{fig:compare_lms_good_explain}
\end{figure}

\noindent In this experiment, we compared the explanation capabilities of MPT explainer with another two state-of-the-art explainable AI methods: LIME~\cite{lime} and SHAP~\cite{lundberg2017unified}, when they are used to explain SVM and BERT's false negative predictions on the adversarial samples (see Fig.~\ref{fig:compare_lms_good_explain}).
Totally, the 500 adversarial samples generated with the {\textbf Configuration 1} are used.
The performance of each explainer is evaluated in terms of the ``good'' explanation metric (see Section~\ref{sec:exp:adv_samples_features_analysis}).

Fig.~\ref{fig:compare_lms_good_explain_a} shows the percent of samples with ``good'' explanations for SVM's prediction behaviors by MPT explainer, SHAP and LIME.
It seems these three explainers explain the model's behavior accurately when the threshold of ``good'' explanation is small.
LIME shows the most stable percentage of ``good'' explanation, as it keeps relatively high percentage of ``good'' explanation in most time even if the threshold is increasing.
MPT explainer shows competitive performance when the threshold is small, but its percentage of ``good'' explanation is decreased when the threshold is increasing.
The percentage of samples with ``good'' explanations by SHAP is deteriorated quickly to zero, as the the threshold of a ``good'' explanation is becoming larger. 
This is because SHAP only explains a small number of features that have the top attribution values.
In summary, we can have a few findings:
(1) Most of the features with high attributed values generated by the three explainers are from the activated features of the adversarial samples.
(2) LIME shows the relatively stable performance, regardless of the thresholds;
(3) MPT explainer has competitive explanation ability with LIME, when the threshold is small;
and (4) SHAP's performance is not stable and deteriorated quickly as the threshold is increasing.
\\

\subsection{Fidelity Test}
\label{sec:exp:fidelity_test}
\noindent We also compare the explanation fidelity between MPT explainer and kernel SHAP, when they are used to explain the PDF malware detectors (e.g., SVM and Random Forest classifiers) for the samples from the PDF malware dataset~\cite{smutz2012malicious}.
% In this experiment, the explanation performance of MPT explainer is compared with Kernel SHAP.
The PDF malware dataset has $4,999$ malicious samples and $5,000$ normal PDF files.
We follow~\cite{guo2018lemna} to use the 135 features, which values have been encoded into binary values. 
The SVM and Random Forest are trained on the training set, and we observe the explanation performance when MPT explainer and SHAP are used to explain the classification of the testing samples.

The fidelity tests include {\it deduction test} and {\it augmentation test}.
The {\bf deduction test} assumes that the AI model's prediction of a manipulated sample, in which a certain number of features with high attribution values (so called ``important'' features) are removed, will towards the opposite class of the original sample, if the explanation (i.e., the attribution values) by the explainer is correct.
The {\bf augmentation test} refers to adding a certain number of features with high attribution values from a malware sample to a benign sample (i.e., the sample in the opposite class) may make the model’s prediction towards the class of malware, if the explanation is correct.
We follow~\cite{guo2018lemna} to evaluate the explainer's performance in the fidelity tests by {\bf positive classification rate (PCR)} that refers to the percentage of samples remains its original class label after the manipulation of deduction or augmentation.
Therefore, the PCR for a ``good'' explainer is as low as possible in a deduction test.
A ``good'' explainer has a high PCR in augmentation.

\begin{figure}
  \centering
  \begin{subfigure}[b]{0.4\textwidth}
    \centering
    \includegraphics[width=\textwidth]{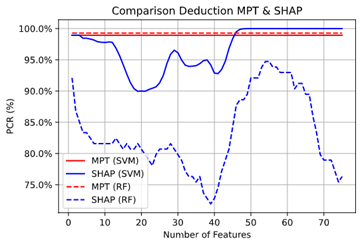}
    \caption{Deduction Test}
    \label{fig:fidelity_test_deduction}
  \end{subfigure}
  \hfill
  \hfill
  \begin{subfigure}[b]{0.4\textwidth}
    \centering
    \includegraphics[width=\textwidth]{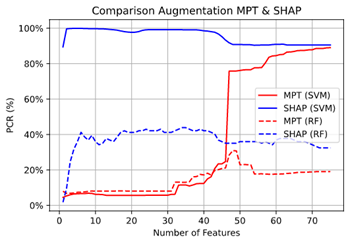}
    \caption{Augmentation Test}
    \label{fig:fidelity_test_augmentation}
  \end{subfigure}
  \caption{\textbf{Fidelity test}. (a) Deduction test: SHAP shows unstable PCRs that fluctuate at the beginning for SVM and increased to neraly 100\% when more than 50 features used that is higher than the PCR of MPT explainer. (b) Augmentation test: MPT explainer has a low PCR when a small number of features used, and increased quickly to close to the PCR of SHAP, which is decreasing, as the number of features is increasing.}
  \label{fig:fidelity_test}
\end{figure}

Fig.~\ref{fig:fidelity_test} shows the fidelity of the MPT explainer and SHAP.
In deduction test (see Fig.~\ref{fig:fidelity_test_deduction}), removing a small number of “important” features (less than 3 features) identified by both SHAP and MPT explainer can reduce the PCR to around 98.92\% for SVM classifier. 
The PCR curve of SHAP then fluctuates sharply when more and more “important” features are used, and it will finally increase to nearly 100\% PCR that is higher than MPT explainer’s PCR, when more than 48 features are used (135 features totally). 
The similar case happens when SHAP explains Random Forest. In comparison, the PCR of the MPT explainer keeps relatively lower than 100\% in a stable way (around 98.92\% PCR).
The unstable PCR curve of SHAP increases the uncertainty of accurately identified “important” features towards the model’s prediction.
We analyzed the explanations generated by SHAP and MPT explainer, and found that SHAP usually only focus on a small amount of features (6\% features for SVM, and 10\% features for Random Forest), which means most of the remaining features are attributed with zeros.
This may cause the problem that if one “important” feature is not identified or an unimportant feature is incorrectly identified in the explanation, the accuracy of the whole explanation will be decreased significantly. 
In comparison, MPT explainer attributes the features by not only considering their individual contributions, but also the contributions made by the dependencies among them. 
Therefore, in both SVM and RF classifiers, MPT explainer attributes more than  60\% features with non-zero values, which forms a large pool of candidate features that have contribution (positive or negative) to the model’s prediction, and these features keeps the stable explanation regardless the number of features used in the deduction tests.

In Fig.~\ref{fig:fidelity_test_augmentation}, it shows that in the augmentation test, adding a small number of “important” features identified by SHAP from the malware samples to a non-malware sample can make the model’s prediction towards malware (i.e., the opposite class of non-malware samples) with a high PCR value ($>$80\%). 
However, as the instability issue for SHAP in the deduction test, as long as more “important” features are used, the PCR of SHAP becomes not stable and intends to decrease significantly. 
The MPT explainer shows a low PCR when a small number of features are used, and a rapid growth of the PCR as the increasing of the “important” features. 
And finally the PCR of MPT explainer is getting close to that of SHAP.
We argue that 
(1) SHAP is only able to accurately identify the contributions of a small number of features, and therefore, the explanation is not complete and the fluctuating PCR curves along the increasing number of features implies the remaining features that are attributed with low or zero attribution values by SHAP can also affect the model’s behavior significantly and likely to mean inaccurate attributions. 
(2) MPT explainer, however, explains the model’s prediction on a sample by attributing more features’ contribution/importance, and the rapid growth of PCR curves in the augmentation test implies that the attribution values for the remaining features by MPT explainer are still accurate.
This is because MPT explainer takes care of both the individual features’ contribution to the model’s prediction and the dependencies among features.

Therefore, we suggest 
(1) MPT explainer can be used to generate a more complete explanation of a model’s behavior on a given sample. Users can understand the AI model's behavior through the contributions made by most of the features in a sample.
(2) SHAP’s explanation on a model’s behavior usually focuses on the contribution made by a small number of features. Therefore, it may be a good practice to use SHAP to have a compact explanation (abstract) of a model's prediction on a given sample.

% \begin{figure*}
%     \scalebox{.40}{\includegraphics[left]{./fig-explain-bert-compare-lms.png}}
%     \caption{\textbf{Comparison: ``good'' explanation} }
%     \label{fig:adv_samples_distribution}
% \end{figure*}

\subsection{Running Time Performance}
The average running time that MPT explainer explains the Android malware detection by SVM for a single data sample is around $15.44$ seconds, and $15.20$ seconds for BERT, respectively.
% The average running time of MPT explainer on a single data instance of Android malware apps features are around $15.44$ seconds for the predictions by SVM and $15.20$ seconds for the predictions by BERT, respectively.
The explanation was running through an un-optimized Python code on a PC with GeForce RTX 2070 GPU, 3.60 GHz $\times$ 8 Intel i7-9700K CPU and 62.8 GB memory.
Most of the time were spent on the prediction process by the classifiers over the hundreds of perturbed data.

\section{Conclusion}
\label{sec_conclusion}
In this article, we presented a novel explainable AI method that addresses the problem of features attribution for machine learning classifiers used in Android malware detection.
This method is inspired by the modern portfolio theory (MPT) that minimizes the variance of the association of the prediction scores changes and the attribution values.
By this way, a higher value of a feature's attribution implies that the small change of this feature will cause a non-trivial change of the model's prediction score.
The effectiveness of the proposed method is assessed by three experiments.
% The first experiment proves that the features with high attribution values identified by MPT Explainer are distinctive.
% It only requires a small number of such features to distinguish a sample from others in terms of the malware families.
The first experiment presents a comprehensive analysis on the capacity of features attribution by the MPT explainer for the adversarial samples, where the malware samples are camouflaged as the benign apps samples and can evade the detection by the SVM and BERT.
In the second experiment, we compare the explanation capacity between the MPT explainer and the state-of-the-arts methods, e.g., SHAP and LIME. 
The results for these two experiments prove thee MPT explainer is helpful for the security analysts to find out the reasons that the classifiers are fooled by the adversarial samples, such that a machine learning model for malware detection that is more resistant against such attacks can be developed.
The third experiment is to test the explanation fidelity by the MPT explainer, where the comparison with SHAP shows the MPT explainer can explain the model's behavior in a high fidelity.
These results hold a promise that the proposed model explanation method is helpful for both AI and cyber security practitioners.
\bibliography{mybibfile}
\end{document}